Extragalactic cosmic ray sources

with very small contribution in the particle flux on the Earth and their study

A.V. Uryson

P.N. Lebedev Physical Institute of the Russian Academy of Sciences, Moscow 119991

**Abstract.** Existence of extragalactic ultra-high energy cosmic ray sources giving a very small particle flux on the Earth is considered. While the majority of cosmic rays is emitted by a source class providing the flux observed on the Earth, this small part of the particle flux is assumed to be accelerated in an additional sub-class of sources of cosmic rays. As an illustration of this sub-class of sources we discuss accretion discs around supermassive black holes where particles are accelerated by electric fields. Due to acceleration mechanism particle injection spectra are assumed to be hard. In this case cosmic ray flux on the Earth is too low for detection. But propagating particles produce in space a noticeable flux of diffuse gamma ray emission and neutrinos. It should be accounted for when analyzing other source models and dark matter models. At energies $E > 10^{19}$ eV cascade neutrino spectra depends on cosmic ray injection spectra. It is proposed to study cosmic ray sources under consideration using data on gamma ray and neutrino emission along with cosmic ray data.

**Introduction**. Cosmic rays (CRs) at ultra-high energies (UHE) - $E>4\times10^{19}$ eV - seems to be accelerated in extragalactic point-like sources, but still they are not revealed. Source identification by particle arrival directions was not effective mainly due to two reasons: errors in CR arrivals of $\sim 1^0$ that are too large to select a source among astrophysical objects falling in the error-box around arrival direction, and to CR deflection in extragalactic magnetic fields which are studied insufficiently at present.

Recently some characteristics of UHECR sources have been obtained exploring cosmological evolution of astrophysical objects where CRs seemingly can be accelerated to UHE, and varying values of injection spectral index $a$ (injection spectra being $\propto E^{-a}$) [1-3]. As a result, evolution parameters of possible sources together with injection spectral indices have been determined that provide excellent fits to energy spectra measured at ground-based arrays the Pierre Auger Observatory (PAO) and the Telescope Array (TA).

Not only a good fit to the spectra measured have been given in [1], [2], [3].

Since UHECRs propagating in space initiate electromagnetic cascades [4], [5] they contribute to the extragalactic isotropic diffuse gamma ray background IGRB. Therefore the intensity of cascade gamma ray emission $I_{cascade\ \gamma}$ should satisfy the condition:

$$I_{cascade\ \gamma} < IGRB - I_{unresolved\ sources}, \qquad (1)$$

where $I_{unresolved\ sources}$ is the intensity of individual unresolved gamma ray sources that is a part of IGRB. The intensity IGRB is measured by the gamma ray telescope Fermi LAT onboard the cosmic observatory Fermi [6], $I_{unresolved\ sources}$ is estimated theoretically [7]. In [1], [2], [3] the intensity $I_{cascade\ \gamma}$ has been calculated in models which fit the measured UHECR spectrum. In this way source parameters have been selected with which condition (1) is satisfied: injection proton spectral index equals to $a$=2.1, 2.2, 2.6 [2], depending on the cosmological evolution of possible sources.

A further step towards the revealing of UHECR sources has been taken in the paper [8], where a minimal model is suggested using a single source class and describing in a unified way not only CR spectra and the CR contribution to IGRB, but the astrophysical neutrino flux from the neutrino observatory IceCube [9] along with the data on the CR composition from the PAO (mean shower maximum depth and its width [10]). (In the model [8] the CR composition and the CR flux are described in the whole energy range above $10^{17}$ eV. The behavior of intensity and the spectral shape of high-energy gamma rays and neutrinos have been discussed in [11].)

Here we consider a possible sub-class of UHECR sources with injection spectra $\propto E^{-a}$ harder than those above. Being inspired by [12] we suppose that hard spectra can be formed when particles are accelerated by electric fields in accretion disks around supermassive black holes (SMBH). Spectral indices above are typical for particle acceleration on shock fronts [13], that can occur, for example, in AGN jets (see e.g. [14]).

Computing of CR fluxes and cascade gamma ray emission, we show that UHECR sources with hard injection spectra give negligible particle fluxes near the Earth, but CRs produce a noticeable diffuse gamma ray flux in intergalactic space satisfying condition (1).

Neutrino fluxes are also generated during UHECR propagation. That is why one more condition on CR models arises: cascade neutrino intensity $I_{cascade\ \nu}$ should be less than the intensity of astrophysics neutrino measured $I_{\nu\ measured}$,

$$I_{cascade\ \nu} < I_{\nu\ measured} \qquad (2)$$

Neutrino flux is obtained, as mentioned above, at the neutrino observatory IceCube, and also at the PAO.

At the IceCube the cosmic neutrino flux has been measured in the energy range of about $(10^6 - 10^{11})$ GeV [9]. At the PAO the tau-neutrino flux is obtained in the range of approximately ($10^{17} - 10^{20}$) eV [15]. In the model suggested cascade neutrino intensity satisfies condition (2) both in the IceCube and in the PAO energy range.

Thus we conclude that possibly extragalactic UHECR sources exist giving insignificant CR flux on the Earth. Characteristics of these sources can be studied measuring extragalactic diffuse gamma ray and neutrino background together with UHECR data.

The computations of particle propagation in the space were performed with the TransportCR code [16].

**The model**. We assume that UHECRs are accelerated in SMBHs, the maximal particle energy $E_{max}$ depending linearly on SMBH mass $M$ [12]: $E_{max}=10^{20}$, $4\times10^{20}$, $4\times10^{21}$ eV for $M=2.5\times10^6 M_\odot$, $10^7 M_\odot$, $10^8 M_\odot$ respectively, where $M_\odot$ is the solar mass. We consider SMBHs with these masses with the ratio $2.5\times10^6 M_\odot : 10^7 M_\odot : 10^8 M_\odot = 0.313:0.432:0.254$ derived from the local SMBH function [17]. The fraction of SMBHs with masses larger than $10^8 M_\odot$ is small comparing with values above [17]. SMBHs with masses smaller than $2.5\times10^6 M_\odot$ produce CRs at energies less than $10^{20}$ eV and their cascade emission is much lower than that of CRs at higher energies [18]. Thus CRs from SMBHs with masses lower than $2.5\times10^6 M_\odot$ and higher than $10^8 M_\odot$ are not accounted for.

The SMBH evolution is unclear and therefore we consider the evolution scenario ([19], see also [16]) of powerful active galactic nuclei - Blue Lacertae objects.

The injection spectra in sources assumed to be exponential, $\propto E^{-\alpha}$, with the spectral index equals to $a=2.2, 1.8, 1, 0.5, 0$ where 0 corresponds to equiprobable generation of particles at any UHE.

We assume that UHECRs consist of protons.

Protons lose energy in synchrotron and curvature radiation in magnetic fields in their path from acceleration region. The problem of energy losses of escaping CRs is discussed in [14]: approximately 0.4 % of accelerated CRs flies away losing insignificant part of their energy.

In intergalactic space UHECRs interact with microwave and radio emissions mainly in reactions $p+\gamma_{rel} \to p+\pi^0$, $p+\gamma_{rel} \to n+\pi^+$. Pions decay $\pi^0 \to \gamma+\gamma$, $\pi^+ \to \mu^+ +\nu_\mu$ giving rise to gamma-quanta and muons, and muons decay $\mu^+ \to e^+ +\nu_e+\bar{\nu}_\mu$ giving rise to positrons and neutrinos. Gamma-quanta and positrons generate electromagnetic cascades in the reactions with cosmic microwave emission and extragalactic background light: $\gamma+\gamma_b \to e^+ + e^-$ (pair production) and $e+\gamma_b \to e'+\gamma'$ (inverse Compton effect).

The extragalactic background emissions are considered as follows. The cosmic microwave background radiation has Planck energy distribution with the mean value $\varepsilon_r=6.7\times10^{-4}$ eV. The mean photon density is $n_r=400$ cm$^{-3}$. The extragalactic background light characteristics are taken from [20]. To describe the background radio emission, we use the model of the luminosity evolution for radio galaxies [21].

Magnetic fields in intergalactic space influence on cascades due to electron synchrotron emission. Fields being of $10^{-9}$ –$10^{-8}$ G and lower the cascade electrons lose energy insignificantly [22]. Though the magnetic field in intergalactic space is apparently nonuniform [23], [24], [25], we suppose that regions with fields higher than above occupy seemingly a small part of extragalactic space, and so extragalactic magnetic fields do not break cascades.

In these assumptions spectra of protons, gamma rays, and neutrinos near the Earth were calculated with the TransportCR code [16].

**Results.** The calculated UHECR energy spectra along with the spectra obtained by the PAO [26] and the TA [27] are shown in Fig. 1. The model spectra are normalized to the PAO spectrum at the energy of $10^{19.5}$ eV ($3.16 \times 10^{19}$ eV). The model CR spectra are lower than the spectra measured by several orders of magnitude (except the point of normalization and the point of about $10^{19.45}$ eV). In addition, the calculated spectra differ greatly in shape from the spectra measured. In the model CRs at energies about $4 \times 10^{21}$ eV fall on the Earth, but their flux is too low to be detected.

We proceed now to the intensity of gamma ray emission that UHECRs initiate in extragalactic space.

Spectra of the cascade gamma ray emission are virtually independent on the initial CR spectrum [25, 28] and here we analyze only the intensity of the cascade emission without discussing the spectra. Namely, we compare model integral intensity of the cascade emission with Fermi LAT data in the range $E>50$ GeV (as it was done in [2]). The reason is that in this range the contribution of discrete unresolved gamma ray sources is estimated in [7]**.**

The integral intensity of the cascade emission $I_{\text{cascade } \gamma}$ ($E>50$ GeV) is found from the differential intensity, which is calculated with the TransportCR code. For the set of spectral index values $a$=2.2, 1.8, 1, 0.5, 0 the cascade integral intensity is: $I_{\text{cascade } \gamma}(E>50 \text{ GeV}) \approx (1.1\text{-}1.6) \times 10^{-10}$ (cm$^{-2}$ s$^{-1}$ sr$^{-1}$).

Now we will compare the model intensity of the cascade gamma ray emission with the Fermi LAT data. Extragalactic isotropic diffuse gamma ray background IGRB measured by Fermi LAT is [6]:

$$\text{IGRB } (E>50 \text{ GeV}) = 1.325 \times 10^{-9} \text{ (cm}^{-2}\text{ s}^{-1}\text{ sr}^{-1}). \qquad (3)$$

This value includes the emission from individual unresolved gamma ray sources. Their contribution to the IGRB at energies $E>50$ GeV equals to 86 ($-14$,$+16$)% [7]. Subtracting from the IGRB the unresolved source contribution of 86%, we obtain

$$\text{IGRB}_{\text{without unresolved sources}}(E>50 \text{ GeV}) = 1.855 \times 10^{-10} \text{ (cm}^{-2}\text{ s}^{-1}\text{ sr}^{-1}). \quad (4)$$

The model cascade gamma ray emission $I_{\text{cascade }\gamma}(E>50 \text{ GeV})$ is less than the value (4) for all spectral indices considered, contributing to the IGRB $_{\text{without unresolved sources}}(E>50 \text{ GeV})$ about 10%. Thus the model under consideration satisfies condition (1).

Neutrinos are also generated during UHECR propagation, whereupon condition (2) on CR models arises. Model neutrino fluxes together with those measured at the IceCube [9] and at the PAO [15] are shown in Fig. 2, the neutrino flux from the IceCube being the strongest limitation. Calculated fluxes are lower than measured ones, so our model satisfies condition (2).

Figure 2 shows that the cascade neutrino spectrum at the energies $E>10^{19}$ eV depends on the injection CR spectrum. This energy range is available for measurement. Thus it is possible to apply neutrino data for analyzing the model under consideration.

**Discussion.** Sources considered in the model obeys constraints arising from gamma ray and neutrino fluxes measured. At the same time these sources contribute negligibly to the CR flux on the Earth. Thus another sources provide the majority of UHECRs, which in turn generate cascade gamma rays and neutrinos when propagate in space. Now we consider if our model is consistent with the minimal model [8], describing the UHECR majority, i.e. data on CRs together with cosmic background emission.

In our model cascade gamma rays contribute about 8-12% to the diffuse isotropic gamma ray emission from Fermi LAT. Thus the model leaves room for gamma rays initiated by UHECR majority.

Neutrino fluxes in the model is also lower than those measured. However in the energy range of $10^{18.7}$-$10^{19}$ eV protons with injection spectral indices $a=0, 0.5$ produce neutrino fluxes which evidently leave not enough room for neutrinos obtained in [8].

The intensity of secondary gamma rays and neutrinos in the Fig. 2 has been obtained normalizing the model CR flux to PAO data at $10^{19.5}$ eV. Normalization of CR fluxes at the energy of approximately $10^{19.45}$ eV reduces them by ≈20%. Then model fluxes of cascade gamma rays and neutrinos also decreases: in the range of $10^{18.7}$-$10^{19}$ eV the model neutrino flux makes up ≈70% of the flux from the IceCube at $a=0$ and ≈50% at $a=0.5$. This is also apparently too high to be consistent with the minimal model [8]. Therefore we conclude that proton injection spectra with indices $a=0, 0.5$ are formed with lower efficiency than it is suggested in our model, or excluded. Proton injection spectra with indices $a>0.5$ are consistent with the minimal model [8].

The normalization above satisfies the only condition: the model particle flux does not exceed the measured one. But it is unknown how much less can it be. Because of this gamma ray and neutrino fluxes obtained in the model are upper limits for cascade emission.

In the model we do not account for proton interactions with IR photons and gas in gas-dust torus surrounding the central part of AGN. Accounting for particle interactions in torus, production of secondary gamma rays and neutrinos increases somewhat [8].

**Conclusion.** Possibly extragalactic UHECR sources with hard injection spectra $\propto E^{-a}$, $a \leq 2.2$ exist. Hard spectra can be formed for instance when particles are accelerated by electric fields in SMBH accretion disks, the maximal particle energy depending on SMBH mass. This acceleration process was analyzed in [12].

We consider SMBHs with masses $M=2.5\times10^6 M_\odot$, $10^7 M_\odot$, $10^8 M_\odot$. The fraction of SMBHs with larger masses is much less [17]. SMBHs with smaller masses produce CRs at energies less than $10^{20}$ eV and their cascade emission is much lower than that of CRs at higher energies [18]. So CRs from them are not accounted for.

Thus any AGN containing the SMBH with a mass above can be a hard UHECR source.

On the Earth the UHECR flux from sources discussed is too low for detection even with giant ground-based arrays. But these UHECRs produce in the space noticeable fluxes of diffuse gamma rays and neutrinos, which satisfy conditions arising from diffuse gamma ray and neutrino fluxes measured by Fermi LAT [6] and IceCube [9].

At the same time the majority of UHECRs is accelerated in another sources or processes and these UHECRs generate in turn cascade gamma rays and neutrinos when propagate in space. The suggested sub-class of sources leaves room for the model emission obtained in the minimal model [8], describing the data on CRs together with cosmic background emission.

The way to investigate the suggested sub-class of sources is to study extragalactic diffuse gamma ray and neutrino emission.

Analyzing diffuse gamma ray emission the contribution from individual unresolved gamma ray sources should be extracted. At present the latter is determined with large percentage error about 15% [7]. To improve it gamma ray telescopes with better angular resolution than that of Fermi LAT ($0.05^0$ at energies above 100 GeV) are required. The comparison of parameters of current and planned gamma-telescopes is presented in [29]. Currently the gamma ray telescope MAST proposed in [29] has the best characteristics at the energies above 20 GeV: depending on the energy its angular resolution is 3-10 times higher than that of the Fermi LAT. The gamma ray telescope GAMMA-400 is also suitable as its angular resolution is of approximately $0.01$-$0.02^0$ at the energy of 100 GeV. It is planned to launch in 2025 [30].

Future neutrino observatories with parameters better than those of IceCbue are listed and discussed in [9] and references therein.

Not only the study of hard UHECR sources considered is of interest. The contribution to the diffuse gamma ray and neutrino emission by UHECRs discussed should be accounted for analyzing cascade emission in any other source models. Also the addition part of diffuse gamma ray emission should be taken into account analyzing dark matter models.

**Acknowledgment.** The author thanks O. Kalashev for discussion of the code TransportCR along with the TA data and M. Zelnikov for discussion of processes in SMBHs. The author also thanks the referee for remarks.

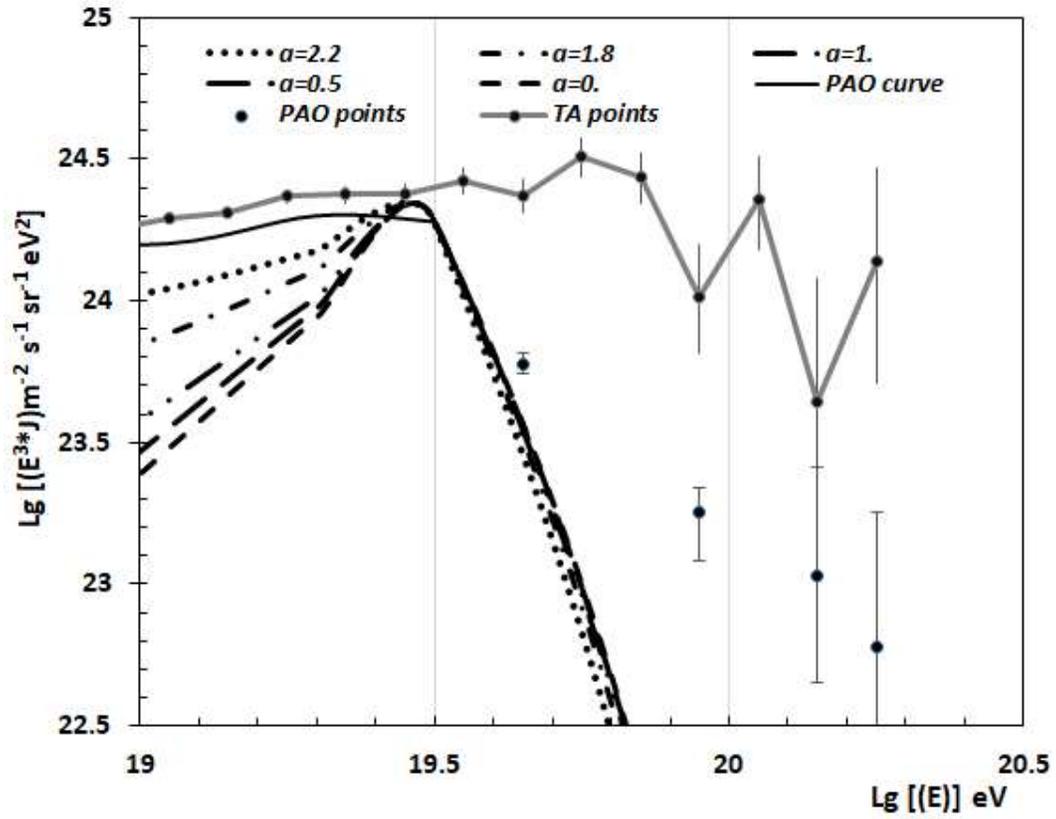

FIG.1. UHECR energy spectra measured on the PAO and the TA, and UHECR spectra on the Earth calculated for injection spectra with various spectral indices $a$ (see the legend). Model spectra are normalized to the PAO data at the energy of $10^{19.5}$ eV ($3.16\times10^{19}$ eV).

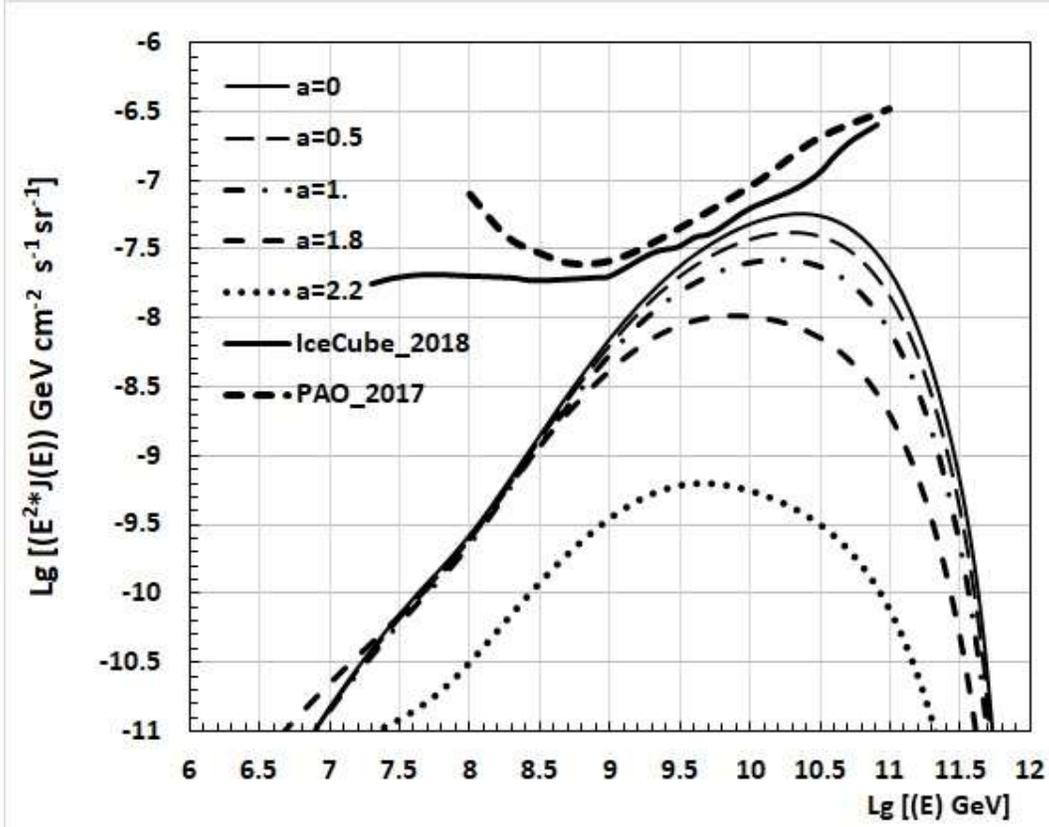

FIG. 2. Model spectra of cascade neutrinos on the Earth calculated for UHECR injection spectra with various spectral indices $a$ (see the legend), the astrophysical neutrino flux from the IceCube [9] (is labeled "IceCube_2018"), and the upper limit on the diffuse flux of tau neutrinos from the PAO [15] (is labeled "PAO_tau").

10